\documentclass[manuscript,review=false,nonacm]{acmart}
\AtBeginDocument{%
  \providecommand\BibTeX{{%
    \normalfont B\kern-0.5em{\scshape i\kern-0.25em b}\kern-0.8em\TeX}}}

\usepackage{todonotes}
\usepackage{tabularx}
\usepackage{varioref}
\usepackage{hyperref}
\usepackage{cleveref}
\usepackage{rotating}
\usepackage{pifont}
\usepackage{multirow}

\renewcommand\footnotetextcopyrightpermission[1]{}
\setcopyright{none}

\begin{document}

\title{Beyond Disinformation: Strategic Misrepresentation across Content, Actors, Processes, and Covertness}


\author{Arttu Malkamäki}
\authornote{Corresponding Author}
\email{arttu.1.malkamaki@aalto.fi}
\orcid{0000-0002-7751-8831}
\affiliation{
  \institution{Aalto University}
  \city{Espoo}
  \country{Finland}
}

\author{Daniel Balinhas}
\email{daniel.balinhasperez@ul.ie}
\orcid{0000-0003-1386-8215}
\affiliation{
  \institution{University of Limerick}
  \city{Limerick}
  \country{Ireland}
}

\author{Letizia Iannucci}
\email{letizia.iannucci@aalto.fi}
\orcid{0000-0002-0635-9933}
\affiliation{
  \institution{Aalto University}
  \city{Espoo}
  \country{Finland}
}

\author{Megan Vine}
\email{megan.vine@ul.ie}
\orcid{0000-0001-5155-0035}
\affiliation{
  \institution{University of Limerick}
  \city{Limerick}
  \country{Ireland}
}

\author{Frederik Temmermans}
\email{frederik.temmermans@vub.be}
\orcid{0000-0002-1986-2183}
\affiliation{
  \institution{Vrije Universiteit Brussel}
  \city{Brussels}
  \country{Belgium}
}
\affiliation{
  \institution{Imec}
  \city{Brussels}
  \country{Belgium}
}

\author{Adrien Coppens}
\email{adrien.coppens@list.lu}
\orcid{0000-0001-7398-2664}
\affiliation{
  \institution{Luxembourg Institute of Science and Technology}
  \city{Esch-sur-Alzette}
  \country{Luxembourg}
}

\author{Nikos Deligiannis}
\email{nikos.deligiannis@vub.be}
\orcid{0000-0001-9300-5860}
\affiliation{
  \institution{Vrije Universiteit Brussel}
  \city{Brussels}
  \country{Belgium}
}
\affiliation{
  \institution{Imec}
  \city{Brussels}
  \country{Belgium}
}

\author{Mikko Kivelä}
\email{mikko.kivela@aalto.fi}
\orcid{0000-0003-2049-1954}
\affiliation{
  \institution{Aalto University}
  \city{Aalto}
  \country{Finland}
}

\author{Michael Quayle}
\email{mike.quayle@ul.ie}
\orcid{0000-0002-7497-0566}
\affiliation{
  \institution{University of Limerick}
  \city{Limerick}
  \country{Ireland}
}

\author{Onur Varol}
\email{onur.varol@sabanciuniv.edu}
\orcid{0000-0002-3994-6106}
\affiliation{
  \institution{Sabancı University}
  \city{Istanbul}
  \country{Türkiye}
}

\author{Fintan McGee}
\email{fintan.mcgee@list.lu}
\orcid{0000-0001-7398-2664}
\affiliation{
  \institution{Luxembourg Institute of Science and Technology}
  \city{Esch-sur-Alzette}
  \country{Luxembourg}
}


\renewcommand{\shorttitle}{Beyond Disinformation}
\renewcommand{\shortauthors}{Malkamäki et al.}
\definecolor{shamrockgreen}{rgb}{0.0, 0.62, 0.38}
\newcommand{\fmg}[1]{\textcolor{shamrockgreen}{#1}}
\newcommand{\Aalto}[1]{\todo[inline,color=orange]{Aalto: #1}}
\newcommand{\LIST}[1]{\todo[inline,color=teal]{LIST: #1}}
\newcommand{\Sabanci}[1]{\todo[inline,color=cyan]{Sabancı: #1}}
\newcommand{\UL}[1]{\todo[inline,color=green]{UL: #1}}
\newcommand{\VUB}[1]{\todo[inline,color=yellow]{VUB: #1}}
\newcommand{\all}[1]{\todo[inline,color=gray]{All partners: #1}}


\begin{abstract}

This article revisits the widely studied problem of disinformation and related phenomena in online social networks (OSNs) by reframing it as a broader problem of \textit{misrepresentation}. While disinformation is commonly understood as the intentional spread of false content, its meaning is applied inconsistently and often remains narrowly content-focused. This obscures other forms of manipulation, such as coordinated behavior that distorts the visibility, popularity or perceived legitimacy of actors and discourses without altering content itself. We argue that such limitations hinder a coherent and operational understanding of information campaigning in OSNs. To address this, we introduce \textit{strategic misrepresentation} as a unifying concept capturing the interplay between content, actors and processes in shaping collective sensemaking. We formalize this concept through a four-dimensional framework encompassing content distortion, actor distortion, process distortion and covertness, reflecting how information campaigns unfold in practice and emphasizing observable behavioral signals. Building on this conceptualization, we conduct an integrative survey of state-of-the-art detection techniques across machine learning, network science and visual analytics. By synthesizing these approaches, we demonstrate how they jointly operationalize strategic misrepresentation in a data-driven manner. Our work provides a novel pragmatic foundation for detecting, classifying and evaluating legitimate and illegitimate information campaigns within and across OSNs.

\end{abstract}






\maketitle

\newpage

\section{Introduction}
\label{sec:introduction}

For information to be meaningful, it must be subject to refutation, wrote Karl Popper \cite{popper_logic_1959}. Popper's criterion, however, was always more an epistemic ideal than an accurate description of how information acquires meaning. In practice, meaning is emergent, arising from collective sensemaking, a process referring to individuals negotiating over the appropriateness of certain attitudes and beliefs to determine what a collective regards as ``truth'' \cite{quayle_social_2025, hayward_problem_2025}.

Throughout the last century, governments, publishers and journalists exercised extensive control over information through mechanisms such as gate-keeping, agenda setting, framing, priming and balance \cite{white_gate_1950, mccombs_agenda-setting_1972, iyengar_is_1991, pan_priming_1997, boykoff_balance_2004}. Such control extended to shaping the perception concerning the popularity and legitimacy of certain actors and discourses, informing audiences not only about the substantive content but also imbuing it with meaning \cite{schmidt_discursive_2008}.  Today, while institutional media power still persists,  the advent of online platforms, and especially online social networks, has rewired these processes, and control over information largely rests with the groups of users engaging with one another on online platforms, \textit{online social networks} (OSNs), as a principal component of our contemporary hybrid media system \cite{chadwick_hybrid_2017}.

When we engage with OSNs, the sheer scale and speed of communication, the affordances of different platforms, together with algorithmic and social (rather than institutional) mediation, often narrow our perception to an increasingly monotonous scope \cite{cinelli_echo_2021}. Consequently, the nature of collective sense-making in OSNs differs markedly from the past, rendering the process vulnerable to harmful information campaigning. The largely home-grown influence operation on Twitter during the 2016 U.S. presidential election, the use of Facebook to incite violence against the Rohingya minority in Myanmar and the amplification of Chinese government content on Douyin illustrate how OSNs afford novel opportunities to distort the social construction of meaning \cite{howard_algorithms_2018, lu_pervasive_2022, schissler_beyond_2025}.

Recently, the term \textit{disinformation} has come to dominate both scholarly and policy debates over illegitimate information campaigning in OSNs. It is commonly distinguished from \textit{misinformation} by denoting the intentional dissemination of veridically false information, such as the fabrication of low-quality news that mimics legitimate journalism in form but not in process or intent \cite{lazer_science_2018}. While the concept of information as a \textit{carrier of meaning} in principle stretches from information about somebody or something to representation by a particular sequence or setting \cite{capurro_concept_2003}, the concept of disinformation is applied inconsistently across research communities stemming from different methodological paradigms and often confined to its content-focused scope \cite{kapantai_systematic_2021, broda_misinformation_2024}. While broader interpretations include actors and processes, such as coordinated behavior, incentivized agents or social bots, these are typically treated as mechanisms surrounding disinformation rather than as part of the phenomenon itself. This creates a conceptual ambiguity, where it becomes difficult to distinguish between disinformation and the means through which it is produced and propagated, and where a primary focus on content risks overlooking manipulation that does not rely on false information. A hallmark of information campaigning in OSNs has indeed been the manipulation of social structure. By engaging with authentic users or artificially exaggerating the visibility of certain actors or discourses, coordinated actors might distort perceptions of popularity and consensus without altering content per se \cite{datta_extracting_2019, varol_online_2017, shao2018spread, keller_political_2020}.

To address these limitations, we urge a shift in focus from disinformation to \textit{misrepresentation}, a broader concept that explicitly captures not only distortions of content but also those of actors and processes that shape users' perceived social embeddedness. This perspective treats content, actors and processes as equally constitutive parts of information campaigns, including cases where legitimate content is strategically amplified to mislead. By reframing the problem in this way, we provide a more coherent and practically applicable foundation that resolves the ambiguity between what constitutes manipulation and the means through which it operates. Building on this, we define \textit{strategic misrepresentation} and present a pragmatic framework for its analysis and classification that reflects how information campaigns unfold in practice, focusing on observable behavioral signals to validate their presence and infer their nature, as well as on distinguishing strategic from spontaneous misrepresentation.

To supplement our conceptual contribution, we conduct an integrative survey of existing detection techniques, showing how they reinforce one another in operationalizing this concept. Our paper follows an Integrative Review approach \cite{cronin_why_2023}, bringing into dialogue social-behavioral and network science perspectives to generate new insights. As such, it is not a systematic or exhaustive review, but a theory-building effort culminating in a four-dimensional, pragmatic model for detecting information operations consistent with what we conceptualize as ``strategic misrepresentation'' (see Section ~\ref{sec:definitionMisrepresentation}).

In so doing, our work moves beyond prior conceptual work and literature surveys. While Kapantai et al. \cite{kapantai_systematic_2021} examine the literature to develop a conceptual taxonomy of diverse disinformation phenomena that differentiates content-, actor-, and process-level analyses, we operationalize and integrate them for the detection and classification of misrepresentation. Regarding surveys focusing on detection techniques, Guo et al. \cite{guo_future_2020} and Bekoulis et al. \cite{bekoulis_review_2021} concentrate on false textual content while Dan et al. \cite{dan_visual_2021} extend attention to other modalities of false content. Hayawi et al. \cite{hayawi_social_2023} review bot detection techniques using deep learning models. Mannocci et al. \cite{mannocci2024detection} address techniques targeting coordinated online behavior, while Hartwig et al. \cite{hartwig_landscape_2024} survey interventions to counter disinformation, both highlighting the usefulness of accounting for the broader social structure. Beyond conceptual operationalization and analytic integration, our survey leverages technical and theoretical insights from multiple disciplines, namely network science, machine learning, visual analytics and social sciences. These interdisciplinary synergies enable us to advance a pragmatic, four-dimensional model to support bottom-up, data-driven assessment of online misbehavior and strategic misrepresentation which is currently lacking from both research and practice.

In what follows, we formalize the notion of strategic misrepresentation (Section ~\ref{sec:definitionMisrepresentation}) and describe our four-dimensional classification framework (Section ~\ref{sec:classificationMisrepresentation}). We then survey detection techniques at the content (Section ~\ref{sec:contentDimension}), actor (Section ~\ref{sec:actorDimension}) and process levels (Section ~\ref{sec:processDimension}) before outlining a unified approach (Section ~\ref{sec:unifiedApproach}) for integrating all three levels of analysis, and consider visual analytics (Section ~\ref{sec:visualAnalytics}) to put a human in the loop to address misrepresentation. Before concluding (Section ~\ref{sec:conclusion}), we discuss important practicalities (Section ~\ref{sec:discussion}), in particular the role of regulation, access to data, and interventions such as deplatforming.

\section{Strategic misrepresentation online}

\subsection{Definition}
\label{sec:definitionMisrepresentation}

As we chart the contours of strategic misrepresentation, we adopt a pragmatic, bottom-up approach. We therefore begin by suspending the assumption of any single, accurate representation of truth or deception. We instead focus on how meaning is socially constructed and contested from observable practices and their effects, with each individual holding a distinct, inherently subjective representation of reality \cite{knoblauch_common_2016}. Hence what we term a \textit{baseline representation} is not an objective ground truth but a collection of subjective accounts of reality that emerge from compression heuristics (i.e. how humans manage, store and recall complex social information) and collective sensemaking \cite{starbird_disinformation_2019, quayle_social_2025}. Although facts about the physical reality do not change, facts about the social reality do, given they rely on definitions, measurements, interpretations and available evidence. For example, the airborne transmission of SARS-CoV-2 was not a fact early in the pandemic until there was enough epidemiological evidence to convince authorities on how the virus, a fact in the physical reality, spreads \cite{lewis_why_2022}. Put another way, along the lines of John Maynard Keynes, a prominent twentieth century economist: ``When facts change, I change my mind. What do you do, Sir?'' 

\begin{figure}[t]
    \centering
    \captionsetup{justification=centering}
        \includegraphics[width=0.49\textwidth, trim={0pt 220pt 460pt 0pt}, clip]{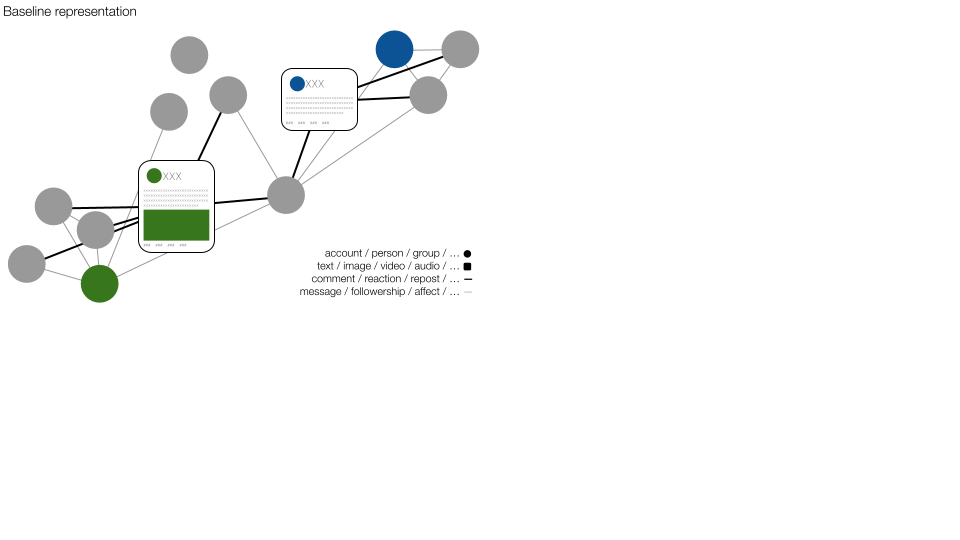}
        \includegraphics[width=0.49\textwidth, trim={0pt 220pt 460pt 0pt}, clip]{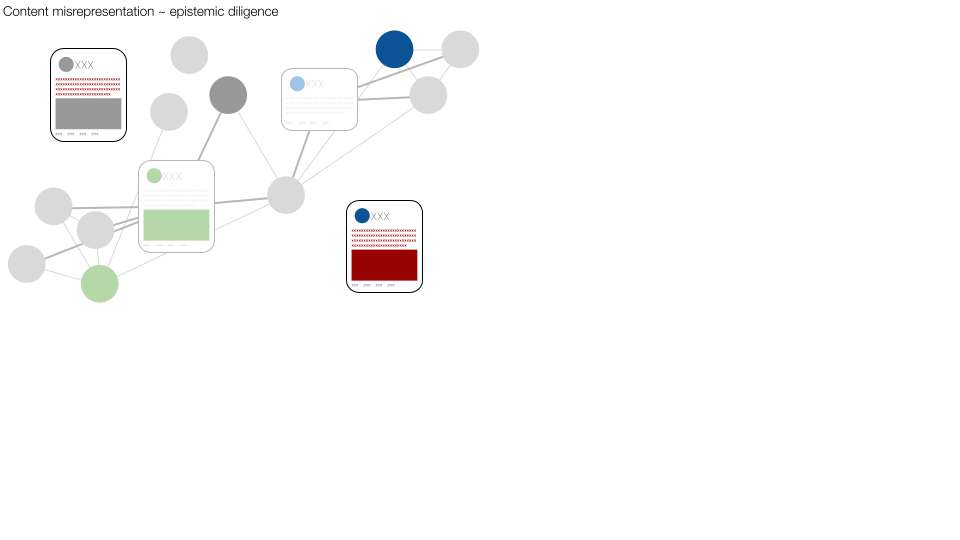}
        \includegraphics[width=0.49\textwidth, trim={0pt 220pt 460pt 0pt}, clip]{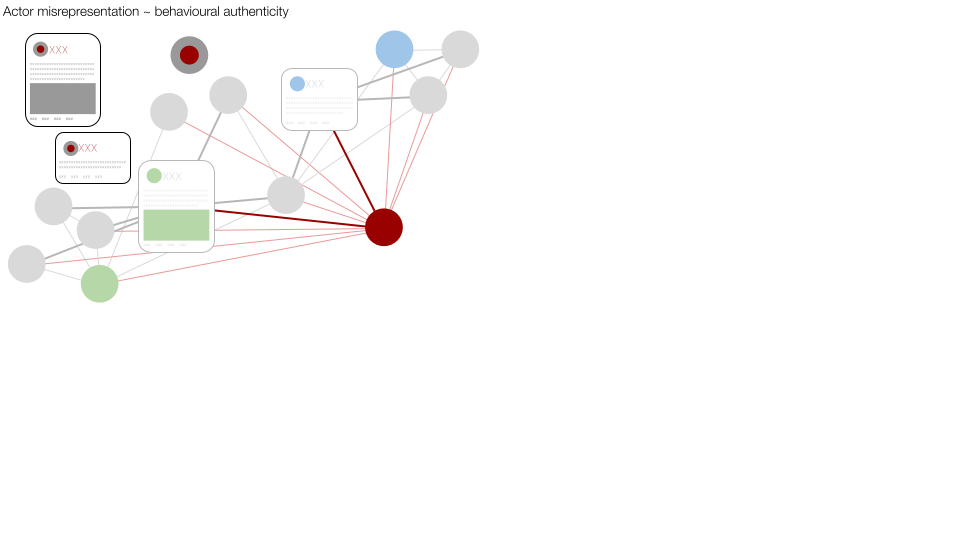}
        \includegraphics[width=0.49\textwidth, trim={0pt 220pt 460pt 0pt}, clip]{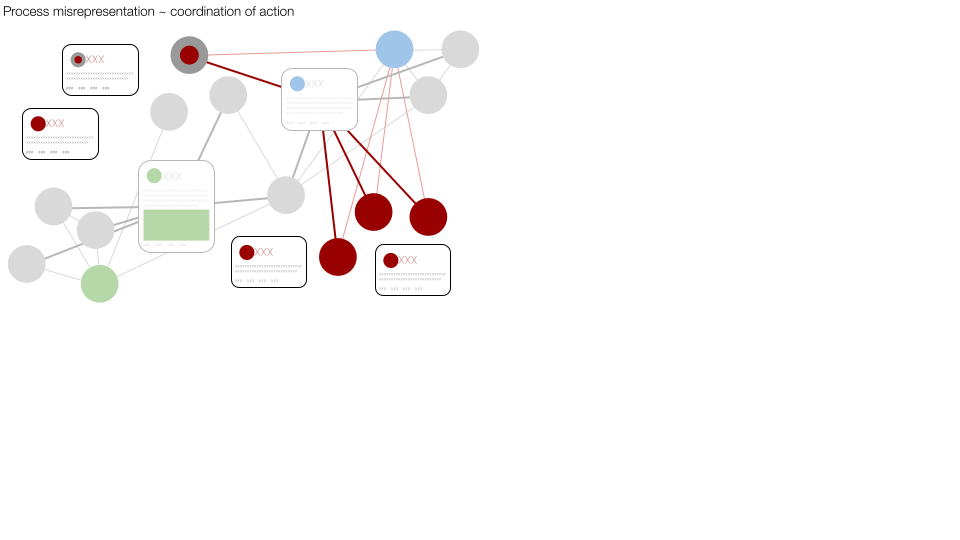}
        \caption{Illustration of a baseline representation online (gray with green and blue to highlight focal users) and its misrepresentation (red against faded baseline underneath).}
        \label{fig:baselineMisrepresentation}
\end{figure}

To render our approach pragmatic, we need empirical data to establish a baseline representation of an OSN. Network science offers a useful data structure, formally representing complex systems such as OSNs as graphs in which the edges capture processual interactions and the two modes of vertices correspond to content items and actor identities, which in the aggregate generate an emergent structure (Figure ~\ref{fig:baselineMisrepresentation} left). Here we consider only a single OSN, but in principle the graph might well extend across several OSNs \cite{Kivela2014}. Assuming the absence of misrepresentation, the graph operationalizes a snapshot of a dynamic baseline representation in line with our methodological standpoint. Misrepresentation then becomes any online manifestation of content, actor or process that ``adds'' to the baseline beyond what would happen without it (Figure ~\ref{fig:baselineMisrepresentation} right). However, we emphasize that each such data structure is an approximation, an incomplete yet tractable mathematical representation of some a slice of the latent social system. 

While some misrepresentation surely is sporadic, spontaneous or benign, it becomes \textit{strategic} when we observe at least one content item, actor or process, or any combination of the three, that deliberately distorts the baseline representation without disclosing it. When it comes to distinguish strategic from sporadic, spontaneous or benign misrepresentations, a major complication is the nature of most OSNs as dramatically asymmetric depictions of the society in terms of mediation, engagement and demography. The vast majority of users act as spectators, while algorithms and influencers, an especially successful minority within the already small subset of content-creating users, bear the greatest responsibility for sustaining the baseline representation \cite{diresta_invisible_2024}. Strategic misrepresentation implies that one or more actors repurpose the exact same actions that sustain the baseline, whether content creation, amplification, recommendation or engagement, to distort it.

Echoing recent scholarly work \cite{condor_public_2006, starbird_disinformation_2019, balinhas_divided_2025}, we therefore formally define strategic misrepresentation online as \textit{an addition to the baseline representation, arising from a collective accomplishment that involves the intentional distortion of content, actors or processes, either within or across OSNs}. Collective accomplishment here implies that the success of strategic misrepresentation relies in part on the joint contribution of algorithms, influencers, and an audience. Since strategic misrepresentation takes hold not from outside the system but from within its ordinary functioning, verifying intent is usually not directly possible. To infer the strategic aspect, we therefore propose accumulating evidence rather than relying on any single indicator.

To identify, first, misrepresentation and then assess the strategic aspect, we propose not only conducting the analyses of content, actors and processes as Kapantai et al. \cite{kapantai_systematic_2021} suggest but also \textit{integrating} them, each yielding input for our four-dimensional classification framework. Stemming from the pragmatist tradition, the framework avoids imposing any \textit{a priori} assumptions about the nature of the phenomena and instead emphasizes observable behavior. It works by prompting the following three questions, which collectively establish whether and what kind of misrepresentation is present: is there epistemically undiligent content (1. Content); are there inauthentic actors (2. Actors); and do interactions among certain actors exhibit coordination (3. Processes). The final evaluative dimension differs somewhat from the three previous dimensions by assessing whether misrepresentation, if any, lacks transparent disclosure (4. Covertness), effectively constraining users' ability to contextualize it and increasing the likelihood that it constitutes strategic misrepresentation. In the following, we further refine and operationalize each dimension.

\textbf{1. Content distortion} constitutes the most basic dimension of our framework and concerns the epistemic quality of online content, including or merging modalities such as text, image, audio, video and hyperlink (i.e. content expansion). Since the degree of inaccuracy to establish such distortion is contingent, we cannot assess it by referring to a ground truth or the ``best available evidence'' thereof. We rather approach content distortion through \textit{epistemic diligence}, focusing on whether claims reflect a genuine effort to ground assertions in evidence, engage with alternative claims and normative assumptions, and remain open to revision \cite{hayward_three_2019, hayward_problem_2025}. Beyond textual claims, such diligence extends to other modalities insofar as images, audio and video make evidential, causal or normative claims through their composition, selection or framing. Epistemic diligence also entails resisting the premature closure of inquiry under institutional or political pressures, and extending critical scrutiny not only to narratives produced outside centers of institutional power, but equally to those advanced through official, institutional channels. Content distortion is present when diligence is demonstrably absent, for example when knowledge claims, supporting data or visual or auditory representations of alternative claims are false, selective or misleading. Most existing detection techniques naturally operationalize epistemic undiligence through observable features of content across modalities and by benchmarking against context-specific corpora, although we emphasize that diligence also implies accumulating evidence across multiple streams. In Section~\ref{sec:contentDimension}, we survey the state of the art in corresponding detection techniques. Given that epistemic undiligence is commonplace in mainstream discourse, observing content distortion alone hardly ever suffices to establish the strategic aspect, although detectable traces of, say, image or video manipulation ought to increase its likelihood.

\textbf{2. Actor distortion} concerns the depiction of actors participating in the OSN. It arises when distortion targets the identity, purpose, or autonomy of users, including impersonation, serial or multiple account creation, and the deployment of human agents or social bots. The distortion here does not reside merely in the presence of particular actors but in how their behavior deviates from that of authentic users as per the baseline representation. We therefore operationalize actor misrepresentation through \textit{behavioral authenticity}, focusing on observable signals such as posting behavior, interaction style, or relational position that disguise the purpose, control or affiliation of the actor. Behavioral authenticity refers to sustained behavioral and interactional patterns (such as posting behavior, reposting pace, frequency, etc.) that could reasonably be produced and/or expected from an autonomous human user. In Section~\ref{sec:actorDimension}, we examine techniques to detect actor distortion.

\textbf{3. Process distortion} addresses whether misrepresentation originates from a deliberate group effort. Such distortion occurs when the principles through which actors organise deviation from the generative processes underlying the baseline representation. Unlike content or actor distortion, process distortion is not attributable to any single entity but emerges from the direct and indirect interactions among actors within or across OSNs. We operationalize it through \textit{coordination}, referring to the presence of systematic, non-random organizing principles that align the actions of multiple actors. Examples include the synchronization of content creation, strong intra-group reinforcement, simultaneous endorsement of certain content, online poll distortion or enduring campaigning to promote a particular discourse. Apparent coordination might well emerge spontaneously, for example when a large number of posts converge on the same topic in response to an event or when civil society actors simultaneously draw attention to an issue without explicit coordination. Establishing process distortion therefore requires distinguishing deliberate orchestration from spontaneous alignment, detection techniques for which we survey in Section~\ref{sec:processDimension}.

\textbf{4. Covertness of misrepresentation.} Covertness constitutes a separate evaluative step of the framework and conditions whether the kind of misrepresentation potentially arising from the preceding analyses corresponds to strategic misrepresentation. Covertness captures whether the distortions across the content, actor or process dimensions lack \textit{transparent disclosure} that would allow users to contextualize the activity. Unlike the three analyses, we do not operationalize covertness as a direct behavioral signal but assess it according to the absence of cues that would normally enable attribution. When disclosure is present, such as in the case of satire, parody, openly partisan advocacy or overt experimentation, misrepresentation might persist but usually does not qualify as strategic. When distortions across one or more dimensions couple with covertness, the activity is substantially more likely to constitute strategic misrepresentation. Worth pointing out is that while covertness may in itself signal the possibility of strategic misrepresentation, there are contexts, such as those in which social actors must conceal their identities, for instance under totalitarian regimes, in which covertness is not only necessary but also legitimate. While the identity may covert, it does not mean that there is a covert misrepresentation, as that would require the anonymous identity to pretend to be somebody they are not.

\subsection{Classification}
\label{sec:classificationMisrepresentation}

The most important contribution of our framework lies in its ability to capture manifold expressions of strategic misrepresentation directly from the data. As such, it can capture both already documented and often-used patterns of online misbehavior through which strategic misrepresentation is frequently enacted, as well as emergent expressions of strategic misrepresentation that do not perfectly fit any ideal type. By this token, in the next section, we use our framework to develop and provide examples of all empirically valid configurations across the four dimensions of our model, thereby showing its potential to capture everyday dynamics of misrepresentation on the ground that may either diverge from ideal-type, previously documented online misbehaviors or constitute a mix of different ideal-type misbehaviors.

To demonstrate the application of our four-dimensional framework, we refer to Table~\ref{tab:classificationMisrepresentation}. The table presents a census of all empirically valid configurations across the four dimensions , each representing an ideal-typical binary category, assigns a qualitative label indicating confidence in a given configuration corresponding to strategic misrepresentation, and provides a minimal illustrative example for each.

\begin{table}[t]
    \centering
    \scriptsize
    \caption{Census of empirically valid configurations across the four dimensions of misrepresentation (CON: Content, ACT: Actors, PRO: Processes, COV: Covertness), with a qualitative assessment of how likely the configuration is to constitute strategic misrepresentation and a minimal illustrative example, including some well-known misbehaviors, for each.}
    \begin{tabularx}{\textwidth}{r c c c c | r p{10cm} }
        \toprule
        \# & CON & ACT & PRO & COV & Strategic & Example activity   \\
        \midrule
        
        
        \textbf{1} & \ding{51} & ? & ? & ? & Unlikely & 
        A conspiracist promoting pseudo-scientific content under own identity. \\

        \textbf{2} & ? & \ding{51} & ? & ? & Unlikely & 
        Users creating secondary accounts to parody a wealthy individual. \\

        \textbf{3} & ? & ? & \ding{51} & ? & Unlikely & 
        Users reacting to a politically divisive news event. \\

        \textbf{4} & \ding{51} & \ding{51} & ? & ? & Unlikely & 
        A sock-puppet posting a daily political satire. \\

        \textbf{5} & \ding{51} & ? & \ding{51} & ? & Unlikely & 
        Users running a political campaign emphasizing achievements and omitting a recent scandal. \\

        \textbf{6} & ? & \ding{51} & \ding{51} & ? & Unlikely & 
        Social bots amplifying diligent public health information \\
        
        \textbf{7} & \ding{51} & \ding{51} & \ding{51} & ? & Unlikely & 
        A cybersecurity agency disclosing a forthcoming information manipulation experiment. \\

        \midrule


        \textbf{8} & \ding{51} & ? & ? & \ding{51} & Possibly &
        An influencer distributes synthetic or manipulated audio, video or images. \\

        \textbf{9} & ? & \ding{51} & ? & \ding{51} & Possibly &
        A user portrays the other side of an argument using one or more additional accounts, cf. sock-puppets \cite{kumar_army_2017, paterson_sock_2024}. \\
        
        - & - & - & - & - & - &
        A user is not who it appears to be due to impersonation or a change in baseline behavior, cf. profile repurposing \cite{elmas_misleading_2023}. \\

        - & - & - & - & - & - &
        A company using a bot to mimic human activity or a real human to promote products or messages. \\

        \textbf{10} & ? & ? & \ding{51} & \ding{51} & Likely &
        A community of users from one domain manipulating an online poll in another domain, cf. brigading \cite{magu_online_2024}. \\

        - & - & - & - & - & - &
        A community of users pretending to misunderstand to frustrate subject experts, cf. trolling \cite{harris_trolls_2023}. \\

        \textbf{11} & \ding{51} & \ding{51} & ? & \ding{51} & Likely &
        As for \#8 but including impersonation via technological means, cf. deepfakes \cite{vaccari_deepfakes_2020, khanjani_how_2021}. \\

        \textbf{12} & \ding{51} & ? & \ding{51} & \ding{51} & Likely &
        Supporters of a political leader targeting a smear campaign against a competitor. \\

        \textbf{13} & ? & \ding{51} & \ding{51} & \ding{51} & Likely &
        Bots carrying out activities such as DDoS attacks or operating as a legion of sock-puppets, cf. botnet \cite{rodriguez-gomez_analysis_2011, bastos_brexit_2019}. \\

        \textbf{14} & \ding{51} & \ding{51} & \ding{51} & \ding{51} & Definitely &
        Disguising a campaign posting undiligent content via bots as a spontaneous grassroots effort, cf. astroturfing \cite{keller_political_2020}. \\
    \bottomrule
    \end{tabularx}
    \label{tab:classificationMisrepresentation}
\end{table}

The table altogether illustrates how our confidence in strategic misrepresentation emerges not from any single dimension in isolation but from their specific combinations. Configuration \#8 (\ding{51} \textbf{?} \textbf{?} \ding{51}) for example captures cases such as a well-being influencer posting inaccurate health claims while obscuring the commercial motivation behind the act \cite{ekinci_dark_2025}. While content distortion and concealment are present, the absence of actor or process distortion limits confidence in strategic intent. Configuration \#10 (\textbf{?} \textbf{?} \ding{51} \ding{51}) instead exemplifies process distortion such as ``brigading'' across online communities, where the distortion lies primarily in the interactions among or across sets of actors rather than in content or actors \cite{datta_extracting_2019}. Configuration \#14 (\ding{51} \ding{51} \ding{51} \ding{51}) in turn reflects an archetypal information campaign such as ``astroturfing'' or ``zombie electioneering'' \cite{keller_political_2020, frost_power_2020}, in which content lacks epistemic diligence, actors' behaviors signal inauthenticity and unusual organizing principles are present, all without transparent disclosure, resulting in high confidence in strategic misrepresentation.

When inferring confidence in an activity constituting strategic misrepresentation, we assign more weight to the third dimension, meaning that the presence of process distortion, in addition to covertness, elevates confidence. Without coordination, covert misrepresentation either at the content- or actor-level, or both (i.e., configuration \#11), might still serve the strategic purpose of sowing confusion among the public.

By enumerating all empirically valid configurations, the framework serves two complementary purposes. First, it provides analytical clarity by disaggregating heterogeneous phenomena. Second, it offers a generalizable routine to reason about uncertainty, allowing scholars and practitioners to qualify claims about strategic misrepresentation even when only partial evidence is available. Rather than treating misrepresentation as a binary outcome, the framework supports comparative assessment across cases, platforms and contexts, while retaining focus on observable behavior. The classification, therefore, functions not as a rigid taxonomy but as a diagnostic tool that guides both empirical inquiry and the integration of diverse detection techniques.

Misrepresentation is not always a result of tactics of common online misbehavior, but the strategic spread of misrepresentations is often enabled by different types of misbehavior. Table~\ref{tab:classificationMisrepresentation} also includes common established online misbehaviors in its examples. It can be seen that covertness is a common characteristic in many of the examples considered. Content distortion, in the form of a clear lack of epistemic diligence, is only inherent to some of them, an aspect which is congruent with our framework's focus on going beyond content alone. This points to the importance of analyzing the interactional features of OSNs holistically (by attending to all four dimensions), beyond epistemic issues, as in many of the ways used to disrupt baseline representations in and of the online social system, deliberate alteration occurs through tactics that target particular system dimensions, usually mixing covertness with actor/identity distortion (such as profile re-purposing) or process distortion (such as brigading and different forms of astroturfing). Our framework precisely allows for this detailed, multidimensional strategy for identifying strategic misrepresentation. 

The framework also suffers from certain limitations. With only binary dimensions, the amount of distortion to meet the criteria in each is inevitably contingent on the baseline. Although we emphasize that the framework assumes connectedness across the analyses across the four dimensions, determining whether a single or a few separate instances of content or actor distortion suffice evidently require further elaboration, for example by introducing additional qualifiers under each dimension. The ratio of distorted content or inauthentic actors under the baseline representation offer an alternative and might prove useful in empirical application. While the framework's unit of analysis is a single campaign, dissecting a campaign of which there could be several in operation at the same time might also pose challenges. Next, we turn our attention to surveying an array of the state of the art in identifying misrepresentation, first, within the content (Section~\ref{sec:contentDimension}), actor (Section~\ref{sec:actorDimension}) and process (Section~\ref{sec:processDimension}) dimensions and then we consider how to unify them, with an ultimate aim of accumulating evidence to detect and classify strategic misrepresentation (Section~\ref{sec:unifiedApproach}).

\section{Detecting strategic misrepresentation}
\label{sec:Detection}
\subsection{Overview}

Detecting strategic misrepresentation calls for accumulating evidence across our four dimensions: content, actor, process and covertness. 
In this section, we provide an overview of the relevant challenges associated techniques from the literature related to identifying strategic misrepresentation. Consistent with an Integrative Review approach, we selectively synthesize the bodies of work most relevant to each dimension of our model. Rather than aiming for systematic or exhaustive coverage, we focus on contributions that directly inform and substantiate the development of a four-dimensional, pragmatic framework for detecting strategic misrepresentation.

Given, however, that our fourth dimension (i.e., covertness) is more qualitative and interpretive in nature, we focus on  techniques to detect misrepresentation in empirical data only across the first three dimensions of our pragmatic scheme. 
Of the first three dimensions content has the largest and most diverse body of work in the literature, and that is reflected here. The actor dimension is extensively studied, however the approaches are based around similar ML techniques with less diversity in approaches. The process dimension is explored less in the literature,as it is a is more recent topic, so therefore  we focus more in detail on the key recent developments.

\subsection{Content dimension}
\label{sec:contentDimension}

Detecting misrepresentation at the content level concerns identifying distortions that are encoded directly in the message itself, independent of who produces it or how it propagates. Much of the existing literature approaches this problem through the narrow lens of false information detection, framing the task primarily as determining whether a piece of content is factually true or false. For example, Guo et al. \cite{guo_future_2020}, Bekoulis et al. \cite{bekoulis_review_2021} and Dan et al. \cite{dan_visual_2021} have recently surveyed techniques to flag false information in OSNs, with an emphasis on content features, multimodal fusion as well as early and explainable detection. Rather than focusing on factual correctness alone, as per our broader misrepresentation lens in Section~\ref{sec:definitionMisrepresentation}, we in turn consider textual or audiovisual content misrepresentative when it fails to meet the standards of epistemic diligence. Yet we also pay attention to research directions and technical standards such as JPEG Trust that go beyond claim verification, for example by explicitly addressing how images might misrepresent not only claims but also actors.

While most detection techniques somewhat inherently establish the lack of epistemic diligence by extracting observable features of content across modalities and by benchmarking against context-specific corpora, which here effectively serve as an evolving baseline representation, we highlight the need to avoid relying on any single technique alone. Rather, we emphasize the need to seek verification across multiple streams, as well as accumulating and updating evidence accordingly. Put another way, the notion of epistemic diligence also applies to the person using our framework to detect and classify strategic misrepresentation. 

\subsubsection{Factual correctness of claims}

Machine learning methods play a central role in the automated detection of low-quality and deceptive content in textual format. These approaches typically rely on a combination of linguistic and contextual features. Linguistic features capture lexical choice, syntax, semantics, domain-specific terminology and degrees of informality, focusing on aspects like word usage, sentence structure and the overall writing style, and which might also signal deceptive intent. Common representations include frequency-based methods such as term frequency \cite{ahmed2017detection, wynne2019content}, term frequency–inverse document frequency \cite{thota2018fake, verma2021welfake}, bag-of-words models \cite{mahir2019detecting} and n-grams \cite{reis2019supervised} as well as embedding-based approaches such as Word2Vec \cite{thota2018fake}, GloVe \cite{bali2019comparative}, and FastText \cite{wani2021evaluating}.

In the task of automated fact checking, natural language processing (NLP) techniques are used to assess the veracity of claims by matching them against evidence sentences. Existing systems are typically trained on small-scale Twitter datasets \cite{Zubiaga2017} or Wikipedia-based corpora \cite{Thorne2018}. Performance on benchmarking tasks such as the Fact Extraction and Verification (FEVER) crowd-sourcing task remains limited, achieving around 74\% FEVER score, and drops substantially when evaluated on real news sources, reaching 49\% predictive performance \cite{Augenstein2019}. Most approaches are monolingual, with limited exceptions such as the transfer-learning-based work of Mohtarami, Glass and Nakov \cite{Mohtarami2019}.

\subsubsection{Distortion by speech and affect}

Textual misrepresentation also operates through psychologically intimidating speech and affect. Hate speech and aggressive sentiment might themselves constitute strategic misrepresentation by distorting the baseline representation, for example by encouraging self-censorship among journalists or researchers working on politically divisive or otherwise controversial topics such as immigration or the ecological breakdown. Sentiment analysis therefore plays an important role in detecting emotional framing that amplifies or disguises an underlying effort to misrepresent the baseline.

Hate speech detection focuses on identifying language that misrepresents people or groups on the basis of attributes such as race, religion or gender, often through de-humanization, stereotyping or coded references. This task is particularly challenging due to linguistic nuance, contextual ambiguity and the deliberate use of sarcasm or memes. Deep learning methods have substantially improved performance, with convolutional and recurrent neural networks capturing local and sequential patterns, and transformer-based models such as BERT \cite{devlin2018bert}, RoBERTa \cite{liu2019roberta} and GPT \cite{brown2020language} establishing state-of-the-art results. Models such as DeepHate \cite{cao2020deephate} integrate contextual embeddings with sentiment and topic information to improve robustness.

Persistent challenges include dataset bias, where explicit slurs are overrepresented relative to subtler forms of abuse, and the rapid evolution of coded language. To address these issues, recent work incorporates multimodal cues and adversarial training strategies to improve generalization to novel forms of affective misrepresentation \cite{mostafazadeh2021improving}.

\subsubsection{Manipulation of audiovisual content}

Misrepresentation can also be encoded in audio, visual or audiovisual content, including images and videos produced or altered using digital cameras, editing software or generative artificial intelligence. Advances in generative models such as Generative Adversarial Networks (GANs) and diffusion models have enabled the creation of highly realistic synthetic visual content, significantly complicating authenticity verification. Such misrepresentation might distort reality, manipulate public opinion and erode trust in visual evidence.

Approaches to the detection of manipulation in audiovisual content broadly divide into reactive and proactive strategies. Reactive approaches analyze the content itself to identify artefacts or inconsistencies, as in deepfake detection. Early methods relied on hand-crafted forensic features, including compression artefacts \cite{agarwal2017photo} and violations of physical scene constraints \cite{obrien2012exposing}, but these techniques often struggled to generalize. More recent work adopts deep learning approaches trained on large corpora of real and manipulated images \cite{wang2019detecting}, including methods based on patch similarity graphs \cite{mayer2020exposing}. Detection of computer-generated imagery has similarly evolved from statistical and color-distribution-based techniques \cite{lalonde2007using} to deep neural approaches \cite{de2018exposing}.

As generative models increasingly suppress obvious artefacts, newer forensic techniques focus on subtler statistical regularities. Examples include co-occurrence matrices \cite{nataraj2019detecting}, color artefact analysis \cite{mccloskey2019detecting} and frequency-domain signatures induced by up-sampling operations \cite{zhang2019detecting}. Transformer-based architectures such as Vision Transformers \cite{dosovitskiy2020image} have also been explored for their ability to capture global spatial and spectral dependencies. Multimodal approaches combining visual, textual and audio signals further improve robustness by checking cross-modal consistency (e.g. features extracted from images and text through foundation models such as the Contrastive Language-Image Pre-training (CLIP) procedure \cite{shang2025semantic, zhang2025knowledge, chen2025multi}.

In contrast to reactive detection, proactive approaches aim to prevent misrepresentation by securely signaling provenance and authenticity information throughout the media life cycle. Standardization plays a key role in this strategy. The Coalition for Content Provenance and Authenticity (C2PA) defines technical specifications for embedding cryptographically protected provenance metadata in digital assets, while JPEG Trust (ISO/IEC 21617) provides a comprehensive international standard addressing provenance, authenticity, integrity and copyright. JPEG Trust is aligned with existing JPEG standards and is compatible with C2PA, extending its functionality by supporting richer provenance signaling and trust evaluation mechanisms.

The first edition of JPEG Trust Part 1, Core Foundation (ISO/IEC 21617-1), was published in January 2025, with a second edition completed in 2026 to align with C2PA 2.3 and add signaling for authorship, ownership and intellectual property rights. The standard supports tamper-evident provenance annotation, extraction and evaluation of trust indicators and privacy and security protections based on ISO/IEC 19566-4. Additional parts addressing trust profiles, watermarking and reference implementations are expected to be completed in 2026 or early 2027.

Together, reactive forensic analysis and proactive provenance standards illustrate complementary approaches to addressing perceptual misrepresentation. While the former seeks to detect manipulation after the fact, the latter aims to embed epistemic diligence directly into the media creation and distribution pipeline, enabling context-dependent assessments of trustworthiness.

\subsubsection{Enhancing content distortion analysis by incorporating actor and process features}
\label{sec:lowQualityNews}

A prominent instance of strategic misrepresentation is low-quality news, often referred to as fake news. It is often characterized by a lack of epistemic diligence and by covert production processes. Importantly, the notion of low quality here refers not only to the surface content but also to the process by which it is created, including the absence of verification, accountability or editorial standards. Such content is frequently designed to mislead, influence public opinion or generate revenue through attention-grabbing narratives and clickbait. Yet textual misrepresentation often becomes detectable by analyzing properties intrinsic to the content itself, including linguistic cues, stylistic markers and semantic inconsistencies, as discussed above.

Beyond linguistic cues, actor- and process-based features are often incorporated to capture patterns associated with the spread of misrepresentative content \cite{castillo2011information, kwon2013prominent, hu2014online, shao2018spread}. Such low-quality news might originate from bots or even from genuine actors (see Section~\ref{sec:actorDimension}, although its widespread dissemination often relies strongly on process distortion, discussed in Section~\ref{sec:processDimension}).

With extracted features, classifiers ranging from traditional methods such as decision trees \cite{ahmed2017detection}, support vector machines \cite{faustini2020fake}, logistic regression \cite{verma2021welfake} and random forests \cite{khan2021benchmark} to deep learning models such as convolutional neural networks (CNNs) \cite{nasir2021fake}, long short-term memory (LSTM) networks \cite{abedalla2019closer}, graph neural networks (GNNs) \cite{bian2020rumor} and transformer-based architectures \cite{abedalla2019closer, wani2021evaluating, khan2021benchmark} have been applied. Graph-based models are particularly effective in the context of OSNs, where propagation patterns and user interactions provide strong signals of misrepresentation \cite{nguyen-etal-2019-fake}. More recently, large language models (LLMs) have also been explored for credibility assessment via prompt-based evaluation \cite{pelrine2023towards}.

\subsection{Actor dimension}
\label{sec:actorDimension}

Strategic misrepresentation is also frequently enabled through the purposeful creation and/or modification of the accounts participating in online social networks (OSNs). The baseline representation can be altered when the identities, purposes or autonomy of users are distorted in ways which deviate from forms of communication that would be expected if actors on social media exhibited behavioral authenticity (see Section \ref{sec:definitionMisrepresentation}).

This section will cover user-based actor distortion and actor distortion produced by the creation of automated accounts. The use of deceptive identities, for instance, involves the covert creation of accounts representing individuals who do not exist and therefore do not reflect—or intentionally conceal—the real user’s offline characteristics (e.g., age, location or nationality, physical appearance, or group membership). Profile repurposing constitutes another form of actor distortion addressed in this section. It refers to the redirection of an existing account (whether originally authentic, inauthentic, or automated) towards a new identity project or (social, political, economic, etc.) function. Furthermore, different types of automated accounts can be created and later repurposed or constitute fake followers, thus enabling the disruption of the baseline representations. Thus, in this Section, we examine techniques to detect actor distortion. Previous works, such as \cite{hayawi_social_2023}, review bot detection techniques using deep learning models. We build on and expand this contribution by putting it into dialogue with studies on actor distortion and by framing bot detection as part of an integrative multidimensional model in which all dimensions are assessed jointly.

\subsubsection{Detection of automated activities of accounts}

There have been important advances in building tools and systems to identify weaponized online systems such as social bots, online trolls, and disinformation websites \cite{ferrara_rise_2016, cresci2020decade}. The arms race between bot creators and detection systems leads to the continuous improvement of these systems for novel challenges \cite{cresci2015fame, Yang2020}. Traditionally, detection systems for automated accounts rely on annotated datasets of different behavior types. Feature engineering has been an important component to capture those automated activities \cite{varol2018feature}. Along these lines, semi-supervised and supervised machine learning approaches have been used to score individual accounts. Botometer is one of the well-known systems for social bot detection on Twitter, and it evolved over the years to address changes in the platform and to capture the novel tactics of the bot creators \cite{varol_online_2017, Sayyadiharikandeh2020}.

There has been a diverse body of research on social bot detection, and a recent review article by Cresci mentions that there were over 230 bot detection systems proposed in the literature between 2010 and 2020 \cite{cresci2020decade}. More recently, Yang et al.conducted a scientometric analysis on the topic \cite{yang2023social}. They analyzed over 900 papers authored by more than 1,600 individuals and all citations received by these papers. Their analysis indicates that detection of automated activities has an impact on scientific disciplines like sociology, political science, and law.

The evidence for automated accounts comes from different disciplines since they have been used to manipulate public discourse, spread disinformation, and pollute organic conversations \cite{ferrara_rise_2016,varol2018deception}, thus potentially distorting baseline representations of some events and of different social actors/groups producing content online. Social bots have been observed to influence the discourse about vaccination \cite{xu2022characterizing, himelein2021bots, seckin2024mechanisms}, climate discussion \cite{daume2023automated, marlow2021bots}, and social protests \cite{uyheng2022bots, stukal2022botter}.

As previously indicated, there are also automated accounts for organizing and disseminating valuable epistemically diligent information, such as making announcements for public services and offering services for social media users. Due to their overt nature these do not indicate strategic misrepresentation. For instance, \citet{lokot2016news} introduces news bots that automate information dissemination of news  and similarly \citet{haustein2016tweets} studied bot accounts that spread research articles on social networks. Emphasis on using social bots for social good can be observed for  mechanisms to introduce public health interventions \cite{deb2018social} and recruiting volunteers to support social protests \cite{flores2016leadwise}.

Of course, models to detect automated activities utilize different approaches to tackle this problem. Supervised learning approaches are more frequently used; however, unsupervised approaches can be more practical for the real-world applications \cite{mazza2019rtbust, mannocci2022mulbot}. This could be because annotated datasets for training may introduce biases and labeled instances may not be a good representation of different characteristics, or it could be due to OSN platforms and observed behaviors evolving with time and models need to adapt to them without requiring new annotated datasets. 

\subsubsection{Deceptive identities}

More recently, generative AI technologies have been used to create realistic fake profile pictures and personas for the creation of accounts. Yang et al. developed a methodology to detect profile pictures that were created using GANs \cite{yang2024characteristics}, estimating that 0.021\% to 0.044\% of Twitter accounts have fake profile pictures. Ricker et al. also analysed nearly 15 million Twitter profile pictures and report that 0.052\% were artificially generated \cite{ricker2024ai}. They also investigated the coordinated activities among them and reported motives including spamming and political amplification campaigns. Another noteworthy study investigates Twitter's information operation dataset and reports fake profile images being active in over 60 different influence operations \cite{george2025forty}. Another strategy applied by bot accounts is using other known users' profile pictures and impersonating them \cite{goga2015doppelganger,zarei2019deep}. 

\subsubsection{Profile repurposing}

In addition to fake visuals, the distortion of accounts also encompasses ``profile repurposing'', through services such as increasing follower counts or boosting engagement of a user's posts. Since creating new accounts is costly, these businesses offer their services from their existing pool of bot accounts, changing their persona and interactions based on campaign goals. Elmas et al. reported large quantities of such accounts when systematically investigating Twitter data on the Internet Webarchives  \cite{elmas2023misleading}. Profile repurposing can involve changes in profile or screen names \cite{hamooni2016url,jain2016dynamics}. This strategy has also been observed in social protests, when users want to signal their position about and issue to broader audience without posting \cite{varol2014evolution}.

\subsubsection{Fake followers}

Automated or compromised accounts can become tools for various misrepresentation campaigns. Some of the earlier examples of the use of bots were the manipulation of the perception of others' accounts by simply increasing their popularity \cite{cresci2015fame,zhang2016discover,aggarwal2018follower}. More recently, an article in the New York Times brought the phenomenon to broader public attention \cite{confessore2018follower}. Similar patterns have been observed on followers of journalist accounts \cite{varol2020journalists}, as well as for manipulations of influencer marketing and the marketplaces for purchasing fake followers \cite{stringhini2013follow,aggarwal2015they}.

Irregularities in the follower properties have been used to identify fake followers \cite{mazza2019rtbust,zouzou2024unsupervised}. Some of these irregular groups can be very large and can go unnoticed by an OSN platform for years. One particular example is the ``star wars'' botnet that posts random quotes from the movie and has a natural-looking number of posts and position in the social network \cite{echeverria2017discovery}. Fake accounts of this nature might look innocent; however, their dormant nature may pose a threat. Such accounts may later be mobilized to amplify content, artificially inflate the popularity of specific accounts, or embed aligned interests within a community, thereby enabling forms of internal manipulation that operate as an ``epistemic Trojan Horse''.

\subsubsection{Actor distortion as one of the bases of process distortion}

Actor distortion, through the covert creation of synthetic accounts, lays the ground (i.e., provides the structure) for synthetic coordination, that is, forms of coordination that can emerge suddenly and unexpectedly when specific social actors attempt to skew public opinion. As we are going to make clear in the next section, strategic misrepresentation is often an orchestrated effort that involves multiple actors collaborating. These efforts often leverage user accounts in their misbehaviors, and this can often include automated accounts, i.e., bots. These can sometimes be used to covertly boost the number of follower accounts to misrepresent the importance of an account (distorting network structure), or also having established a significant presence, be repurposed toward misrepresentation in a coordinated manner.

Indeed, characteristics of the actors and their behaviors can play a role in identifying low-quality news and disinformation (see Section \ref{sec:contentDimension}). The spread of disinformation is often influenced by network dynamics, and understanding these dynamics is crucial for effective detection. Network-based features extract information from the network structure itself, like actor interactions and propagation patterns, which are crucial for understanding how information spreads through networks \cite{bondielli2019survey, nguyen-etal-2019-fake}, thereby detecting misrepresentation attempts like brigading or astroturfing. Similarly, consensus can be modeled as a network property \cite{Salloum2022}, and using network models of consensus allows us to detect group structure and observe its evolution \cite{Dinkelberg2021}.  Furthermore, perceived consensus is a key target of misbehavior, such as misinformation, bot activity, and brigading, which aim at manipulating the perception of widespread (dis)agreement on a specific position \cite{woolley2017}.

Previous work has shown that features like number of followers, the age and gender of users, and news' propagation patterns \cite{Do2019} combined with graphs capturing interactions of social media users or news items \cite{HuuDo2021} can capture coordinated attempts to make particular narratives seem popular. Additional features, such as visual and sentiment-based features that assess the emotional tone of the content, as well as multimodal features that extract information from images or videos and text, can be explored in these network based approaches to assess the authenticity of news \cite{jin2016novel, shu2017fake}.

One problem with the account-centric analysis to identify misbehavior is that the automated behavior online has become more coordinated, multifaceted, and convincing. Following the recent advancements in deep-learning technologies, bot creators can build conversational agents, realistic profile pictures, and fabricated content. However, accounts created by using these technologies exhibit similarities in their content creation, timing or network compositions. These commonalities allow their detection, and the use of such bots as part of coordinated misrepresentation presents opportunities for their identification, which can be used in conjunction with the approaches discussed in Section \ref{sec:processDimension}, and the construction of multilayer network models to address misrepresentation (see Section \ref{sec:unifiedApproach}).

\subsection{Process dimension}
\label{sec:processDimension}

Graphs (or networks) are essential structures used to model and analyze relationships between numerous entities~\cite{newman2018networks}. In the context of OSNs, users can be represented as nodes in a graph, with 2 relationships (such as being friends, interacting with each other, or interacting with the same entity) depicted as edges. The modeling of these interactions makes graphs very suitable for detecting process distortion. As discussed in Section~\ref{sec:definitionMisrepresentation}, process distortion addresses whether misrepresentation originates from a deliberate group effort. This allows for us to expand the focus of identifying strategic misrepresentation to include coordinated behaviour, the key driving factor for process distortion. As identifying coordinated behavior is a more recent focus of the literature, this section covers a smaller range of techniques than the previous sections, but goes into more detail about the fundamental approach.

\subsubsection{Coordinated behavior}

Analyzing coordinated behaviors \cite{keller_political_2020, nizzoli2021coordinated} is crucial because, on one hand, automated behaviors become increasingly sophisticated and harder to distinguish from genuine human interactions, and, on the other hand, information misrepresentation often results from the online activities of both human-operated and automated accounts, with varying distributions \cite{pacheco2021uncovering, nizzoli2021coordinated, cinelli2022coordinated, kirdemir2023towards}. Additionally, it puts emphasis on  identifying the collective strategies of influence campaigns. Studying the online actions of individual accounts in isolation can conceal the inauthentic and coordinated nature of their activities \cite{lynnette2022online}, whereas examining them in the context of other social media users reveals the coordinating group and their common behaviors. In this sense, OSN platforms exhibit a typical characteristic of complex systems, necessitating analysis of the relationships and interactions between their individual components.

Central to detecting inauthentic behavior is the assumption that legitimate users typically demonstrate some level of uniqueness in their online actions. Since a lack of individuality among a group of accounts can indicate coordinated activity \cite{pacheco2021uncovering, pacheco2020unveiling, hristakieva2022spread}, identifying coordinating accounts involves analyzing the timing of their actions, their content sharing practices, and other patterns in their digital footprints. While coordination can be identified by analysis of user interactions, inferring a strategic aspect can often be difficult due to its covert nature -- a prime instance demonstrating the need for our integrative approach.

Both supervised and unsupervised techniques are feasible for misbehavior and coordination detection; however, unsupervised techniques have a clear advantage since they can adopt evolving nature of misbehavior and lack of annotated datasets for training models \cite{pacheco2021uncovering}. The performance of both techniques relies on the representations created for the analysis. Unsupervised learning techniques use these representations for i) learning similarities between entities and ii) grouping similar entities to isolate coordinated groups. To build a detection system, labels for misbehavior and normal activities and vectorized representation are used to train supervised models.

There are also other approaches than pure supervised or unsupervised learning. For example,  Luceri et al. analyzed a data set concerning Russia's Internet Research Agency on social media with a reinforcement learning approach to identify misbehavior at the account level (i.e. coordination) \cite{luceri2020detecting, luceri2024unmasking}. Zhang et al. \cite{zhang2023capturing} propose a model that incorporates a timeseries and content encoder models to classify coordinated cross OSN campaigns on Twitter and Reddit datasets, based on coordination data identified within each of the datasets using either supervised or unsupervised approaches.

\subsubsection{Latent Coordination Networks}

\begin{figure}
    \centering
    \includegraphics[width=0.99\linewidth, trim={0pt 200pt 0pt 0pt}, clip]{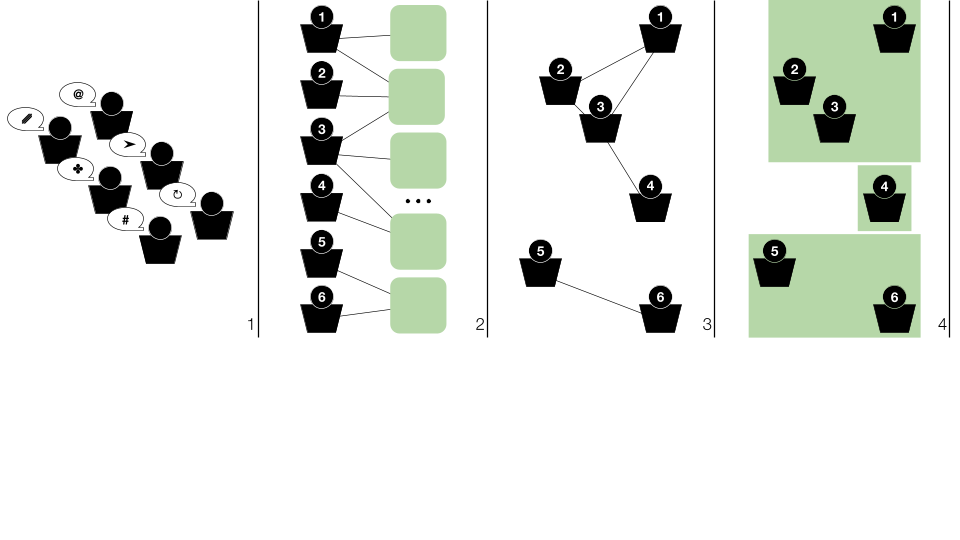}
    \caption{Construction of a Latent Coordination Network (LCN) begins with encoding online user activities into a bipartite graph linking each user to their online actions. Edges can be weighted to reflect the frequency or significance of these actions. The bipartite graph is then projected into a user-to-user network, where edges represent behavioral similarity between the two users. The resulting LCN reveals implicit coordinated groups of users that may not be evident through direct interactions.}
    \label{fig:lcn_construction}
\end{figure}

The challenge of identifying coordinated behavior has prompted the formulation of latent coordination networks (LCNs) \cite{weber2020gang, weber2021amplifying} to better represent such behaviors on social media platforms. In LCNs, an edge between two users signifies shared digital behaviors observed in both nodes' digital footprints with similar patterns \cite{pacheco2021uncovering, pacheco2020unveiling} (Figure~\ref{fig:lcn_construction}). Thus, LCNs can reveal hidden or implicit relationships between users, beyond direct interactions.

Building a latent coordination network begins with constructing a bipartite network that links individual accounts with the set of online activities they have participated in \cite{pacheco2021uncovering}. Bipartite networks are graph structures where nodes belong to two distinct sets \cite{newman2018networks}. Since edges can only connect nodes from opposite sets, bipartite graphs effectively model relationships between different entities, such as social media users and their online engagements. In the bipartite network, edges may be weighted to reflect the strength of these relationships. For the analysis of coordination, edges are typically weighted to emphasize that repeated instances of the same online action provide stronger evidence of coordination than isolated occurrences. Additionally, normalization techniques like IDF (Inverse Document Frequency) can adjust for actions that are more prevalent across the social media platform \cite{pacheco2021uncovering, hristakieva2022spread, luceri2025coordinated}.
 
By applying a selected metric or similarity score over the online behaviors \cite{lynnette2022online, pacheco2021uncovering, pacheco2020unveiling, nizzoli2021coordinated,kirdemir2023towards}, the bipartite network described above can be transformed into a user-to-user latent coordination network. This network usually incorporates undirected, weighted edges that depict potential implicit collaborative dynamics between users. However, a challenge arises in distinguishing edges or weights that result from random chance, which may not accurately represent collaborative intents \cite{linhares2022uncovering, gomes2022networkbackbone}. One strategy to address this challenge involves employing network backbone extraction methods. These methods selectively retain significant signals while filtering out irrelevant noise, thereby enhancing the accuracy of the latent coordination network. Common backbone extraction methods include thresholding, disparity filter, high salience skeletons, spanning trees, or multiscale filtering methods \cite{pacheco2021uncovering, linhares2022uncovering, nizzoli2021coordinated}. Threshold-based approaches, which are the most common, discard weaker or less significant edges, using an arbitrary threshold \cite{pacheco2021uncovering, pacheco2020unveiling, giglietto2020takes, lynnette2022online, kirdemir2023towards}. By doing this, only the nodes having edges that meet or exceed this criterion are further analyzed for potential coordination. Moreover, this process potentially disconnects the network into a number of connected components, that are then separately analyzed.

After filtering out edges that represent noise or irrelevant connections \cite{kirkley_fast_2025, yassin_evaluation_2023}, the refined topological structure of the network and that of its connected components can be systematically analyzed to uncover anomalies or identify clusters. Anomalies within the network (e.g., edges with very high weights) may indicate unusual patterns of interaction or unexpected relationships between users. Topological analysis of the network can reveal essential structural properties such as centrality, community structures, and connectivity patterns. In the context of coordination detection, network centrality measures can help identify the orchestrators of a campaign or influential nodes that have a key role in spreading a narrative. Similarly, community structures -- uncovered via label propagation, modularity maximization, k-cores, or k-cliques \cite{pacheco2021uncovering, linhares2022uncovering, kirdemir2023towards} -- can reveal cluster of users who exhibit a high degree of collaborative behavior, effectively identifying the coordinated groups or botnets \cite{linhares2022uncovering}. Finally, connectivity patterns and network properties, such as degree distribution, clustering coefficient, and shortest path length, illustrate how nodes are linked and how collaboration unfolds within the network. A high clustering coefficient, for example,  suggests the presence of tightly-knit groups frequently engaging in similar activities, while the degree distribution can reveal a hierarchical structure where a few central accounts are responsible for disseminating content to many others \cite{kirdemir2023towards}.

\subsubsection{Temporal filtering of LCNs}

Given the critical role of time for coordination, many approaches to constructing a coordination network begin by establishing a relevant time window over which the LCN is constructed and analyzed \cite{giglietto2020coordinated, giglietto2020takes, magelinski2021synchronized, lynnette2022online, linhares2022uncovering}. Such a time window is meant to capture the temporal closeness of actions that may indicate coordination. In fact, the repeated occurrence of the same action by different entities within this time window can be considered suspicious and indicative of potential coordination \cite{giglietto2020coordinated, magelinski2021synchronized}. For this reason, temporal filtering is a foundational step in many coordination network analyses, ensuring that only temporally relevant interactions are included in the coordination network. As a result, these methods highlight instances where the timing of actions suggests a level of organization or collaboration that goes beyond mere coincidence. A suitable time window for the analysis of coordination can be estimated from the distribution of share times within the dataset, for example by considering the median time taken by the fastest 10\% of the posts to reach 50\% of their total shares \cite{giglietto2020coordinated} or by considering properties of the resulting network, such as its clustering coefficient or average modularity \cite{lynnette2022online}.

It is important to notice that the size of the time window is a crucial parameter in these approaches \cite{giglietto2020coordinated, magelinski2021synchronized, lynnette2022online}. If the time window is too short, the framework might miss coordination events because the actions must occur in a very brief period to be considered coordinated. On the other hand, a wide time window might capture actions that are merely coincidental, leading to identifying synchronicity that is not actually meaningful. Thus, careful selection of the time window size is essential to accurately detect genuine coordination without including coincidental actions \cite{magelinski2021synchronized}. In addition to this, the fixed-window approach has other drawbacks: it requires to continuously slide the window in time to avoid losing up to 50\% of potential connections between users, and produces a number of networks that are separately analyzed. This results in a limited view of the coordination effort and potentially misses significant connections and broader patterns of coordination over time \cite{magelinski2021synchronized, tardelli2024temporal}.

In general, both threshold-based approaches and time window approaches limit the analysis by reducing a complex phenomenon to a binary classification (i.e., coordinated/not coordinated) \cite{hristakieva2022spread, nizzoli2021coordinated}. To overcome this limitation, one approach involves using multiscale filtering methods to retain statistically significant edges, regardless of their scale, and then applying progressively tighter similarity thresholds to these edges \cite{nizzoli2021coordinated}. By repeating community detection with a moving threshold, this method enables the study of how coordinated communities evolve at various levels of coordination and provides a continuous measure of the degree of coordination.

Yet, all the methods mentioned above are limited in that they rely on static analyses of networks that result from aggregating user behaviors over time. As coordination dynamically evolves in time, aggregated networks might conceal important temporal dynamics \cite{tardelli2024temporal}. To address this, Tardelli et al. \cite{tardelli2024temporal} utilize a multiplex temporal network and uncover groups of users exhibiting coordination over time through dynamic community detection.

\section{Unified approach to detecting strategic misrepresentation} 
\label{sec:unifiedApproach}

As discussed in the previous sections, misrepresentation is a multifaceted problem: It comes in the form of misrepresenting content (Section \ref{sec:contentDimension}), actors (Section  \ref{sec:actorDimension}), and distortion of process and the underlying social networks (Section \ref{sec:processDimension}). Further, these categories include rich tapestries of various types of misrepresentations. While most of the methodologies are focused on a particular kind of misrepresentations, for example, detecting low quality news (see section \ref{sec:lowQualityNews}), the different types are not orthogonal. For example, a set of bot accounts with fabricated identities and deepfake profile pictures can be distributing various low quality news articles and other disinformation, and at the same time infiltrating social groups and manipulating recommendation algorithms by manufacturing engagement. 

The multifaceted and clustered nature of misrepresentation calls for a holistic view of the phenomenon. This has two main advantages. First, it can help to understand the motivations, and possibly the intent of the malicious actors and can reveal hints about their identities. Second, it allows for better detection algorithms as one can accumulate evidence from different activities. While methods for hiding individual misrepresentations might get better, for example, bots might have deep fake profile pictures that are hard to detect or realistic distributions of activity times. Still, when hundreds or thousands of accounts are realistic in a very similar way in both contexts, it provides robust evidence for inauthentic and fabricated accounts.

Network approaches can be combined with machine learning approaches to identify misbehaving entities such as social bots, and behavioral indicators (e.g., coordination). At the content-level, anomalous patterns of interaction and amplification of certain types of content within a short time windows are also signatures of misbehavior. The use of ML to identify, characterize, and track the spread of misinformation benefits from a network-aware approach, where we know the wider context of each actor involved and content spread. Graph representation learning approaches provide important avenue to combine machine learning approaches with network analysis as they can be incorporated to build user profiles and learn representations for diffusion patterns. Recently, graph neural network model have been used to identify online information operations \cite{minici2025iohunter}, identification of low quality news spreading patterns \cite{zhu2024propagation,zhang2024heterogeneous}, and early detection of rumor cascades \cite{xia2024lgt,wei2023dgtr,chang2024novel}.

Ultimately for action to be taken to address strategic misrepresentation a human must be involved (even to just approve the findings of various approaches), and must be able to understand complex data that may indicate strategic misrepresentation. Information visualization, i.e., ``the use of interactive visual representations of abstract data to amplify cognition'' \cite{ware2019}, provides a means for this to be done. XAI (eXplainable AI) methods, that is, a set of tools and methods to explain how ML algorithms process data, make decisions and on what (parts of) data are based when making decisions, are important to build trust in ML based systems. XAI methods can also be combined with visualization approaches \cite{yang2019xfake}.

\subsection{Multi-layer approach to understanding strategic misrepresentation}

The multilayer network framework \cite{Kivela2014} offers a way to integrate the content and network structure of OSNs in order to understand, detect, and mitigate misrepresentation in these networks. This is in contrast to understanding misrepresentation as an isolated activity or confined to a single co-activity network. A multilayer approach allows one to combine such networks and understand the complex patterns of misrepresentation and their consequences with meaningful impact. Orchestrated manipulation campaigns, for example, can only be fully understood in the context of different social and knowledge networks \cite{Beskow2019}. 
Similarly, while polarized structures in OSNs can be a natural consequence of democratic processes, they potentially become harmful when they align over multiple topic networks \cite{Chen2021}, and such structures can be susceptible to manipulation \cite{Xia2021}.

A multilayer network approach begins with the creation of a multilayer network representation of the underlying data. The challenge lies in constructing the layers of social interaction and content that are appropriate for the task of modeling the social system. One can approach this layer construction problem from a technical perspective or from a theoretical one.

Technically, one can construct layers of different platforms, different types of interactions, or different co-activities. For example, discussion content such as URLs, hashtags, and other items can be encoded as network layers. The previously described LCN approach for coordinated activities can also be used to define layers. Further, image and video processing will create opportunities to extract content from these media and represent them as co-occurrence networks. Entities in layers can be annotated with the outputs of the previously described ML approaches and this information used to define new layer definitions. Finally, likes, shares, comments, and replies represent different levels of engagement and influence. This approach allows researchers to analyze how both authentic and inauthentic social media actors operate across these networks and how misrepresentation emerges from various interaction types and content contexts.

OSNs are inherently multiplex beyond the technical artifacts, as different topics and narratives exist within the same social network, and users engage in various types of interactions and consume/produce diverse content. For instance, discussions around climate change, immigration, and nationalism might involve overlapping sets of users but distinct content features and engagement patterns. These layers can be identified via manually curated keywords \cite{Chen2021}, topic modeling \cite{salloum2025politics}, or other means such as their diffusion patterns \cite{xia2022limits}.

\subsection{Visualization to understand the unified perspective}
\label{sec:visualAnalytics}

Interactive visualization combined with automated analysis, often referred to as visual analytics \cite{Keim2008} provides an important means by which a human can understand the complex structure of the multiple layers of an OSN (Section~\ref{sec:processDimension}), as well as giving insight into the machine learning techniques that are used to identify content distortion (Section~\ref{sec:contentDimension}) and actor distortion (Section~\ref{sec:actorDimension}). Putting the user in the loop enables them to better understand the context in which  entities are exposed to strategic misrepresentations, as well as the impact on strategic misrepresentation on the wider network, while leveraging their own experience when analyzing the data.

\subsubsection{Content distortion visual analytics}

MisVis is an  online visualization tool that looks at online misinformation at the level of websites \cite{lee2022}. It provides a visual summary of a website's reliability based on the distribution of its hyperlinked websites, showing how many of them are classified as false information (using the classification of \cite{sehgal2021}), and a small network view showing the connections between the websites. MisVis also has an online social media component which focuses on twitter data as means to provide further context. A view is provided that visualizes, for a given target website, the amount for misinformation websites that are also mentioned by users who have shared  that target website. It also shows the number of twitter users mentioning the target website that are bots, according to botometer \cite{Sayyadiharikandeh2020}. In other words, Twitter accounts are used as a means by which the reliability of websites can be judged. ClaimVerif~\cite{ClaimVerif2017} is a text based tool which, given a statement, returns a list of sources along with an associated judgment, including confidence score on whether the claim is true or false. The tool uses little visualization, displaying snippets of text that are evidence related to the tools' judgment. 
ClaimVis~\cite{claimViz2020} is a more recent example of a visual analytics tool that is used to fact check statements in a journalism context. The visualization includes a map of the input transcript, allowing users to identify where fact checked claims occurred. There are a wide range of text visualization techniques that go beyond the text based approached used by the preceding applications, such as text stream visualization e.g., \cite{Dork2010} and topic-based visualizations, as used by \cite{Andreadis2021}. Text sentiment is also a large visualization topic in its own right \cite{Kucher2017}, whereby features can be coded for sentiment (e.g., by color).

\subsubsection{Actor distortion visual analytics}

There are few examples for visual analytics tools focusing primarily on actors. The most prominent example is Verifi2 \cite{verifi2019}. It supports analysis at the source  and account level rather than  the news story level (although it does include analysis of content). It uses analysis of linguistic, network and image features to identify suspicious accounts. It classifies accounts based on their links to other accounts that are in a list of verified and suspicious accounts~\cite{volkova2017}. It also classifies the language of the posts, as well as the images they contain and can be used to better see the difference between news from verified and suspicious sources. Following an evaluation with expert users, the creator of the tool concludes that it is  better to consider a spectrum of trustworthiness rather than a binary classification of real vs suspicious. This is also a motivation of our use of the term misrepresentation. 

\subsubsection{Process distortion visual analytics}

Analysis of graphs is fundamental to understanding the process dimension. Among the freely available general purpose graph visual analytics tools, Gephi \cite{Bastian2009} is one of the most popular and it can be used to quickly visualize social networks, including the results of centrality analyses. Tulip \cite{Auber2018} is targeted toward general graph analysis but can handle graphs at larger scales and custom analytics can be provided using a python interface.
Neither  of these tool are targeted specifically towards multilayer networks in their default configuration. The MuxVis tool provides fundamental visualization for Multilayer Networks \cite{DeDomenico2014} suitable for many domains including the analysis of OSNs. Ghani et al. \cite{Ghani2013} provide an early example of a visual analytics tool specifically for  social networks, which can be considered a multilayer network approach. The authors demonstrate the tool using an authorship-based social network. Where the networks are multilayered, more powerful complex interactions can be leveraged. The multilayered network analysis tool \textit{Detangler} \cite{Renoust2015} features a unique interaction that allows items on a layer to be selected  based on a selection in another layer, using a custom \textit{entanglement} metric. It is a  good example of using the relationships in one layer of a multilayer network to better understand another.

More recent examples can be seen to be targeting the specific example of false information on social media. For example, the web-based visualization platform of \cite{Andreadis2021} provides a front end for an underlying analytics platform. It is focused around geographic views of online social media posts related to the Covid-19 pandemic but does include a node link visualization for community detection based views, as well as word clouds for topic visualization.

Most tools that focus on multilayer networks use \textit{a priori} layer definitions. However, the ability to dynamically  define and simplify layers is very useful for any analysis. Interdonato et al. \cite{Interdonato2020} specify techniques for multilayer network simplification, through the transformation and definition of layers, but supporting this via  visual analytics is  still an outstanding challenge \cite{McGee2021}.

There are also tools with strong visualization components provided by  research institutes that focus on social media that can be adapted to address strategic misrepresentation. The social media lab of Toronto Metropolitan university provides a range of applications that leverage visualization to communicate their message \cite{SOcialMediaLab2025}, including network visualization to understand the exchange of information between telegram channels and heatmap tables to show the spending on political advertising on multiple social media platforms. 

The OsomeNet tool \cite{OSoMeNet2025}, provided by the Observatory on Social Media at Indiana University, uses state of the art network visualizations to better understand the spread of information across multiple platforms. It supports co-occurrence networks (generated from posts containing a specific hashtag) as well as diffusion networks, which visualize how information flows from user to user. OSoMeNet uses Helios-Web \cite{Helios2023} to provide real-time visualization of the underlying dynamic networks.

In the case of coordinated activities, it is not just individual accounts that are of interest but communities of accounts. Within the Verifi2  application a network clustering algorithm is used to group users and specific terms around topics, together in a network view. In terms of coordinated behavior specifically, the Coordiscope tool \cite{coordiscope2025} (also provided by the Observatory on Social Media at Indiana University in addition to \cite{OSoMeNet2025}) detects potential coordination between accounts and visualizes the results as a network.

Visualization can also be leverage to understand graphs at a structural level beyond the low level node and edge definitions. It can be used  to support identification of subgraph structures (motifs) that may characterize misbehavior related to strategic misrepresentation. Cakmak et al. \cite{Cakmak2022} leverage a pixel-based visualization approach to look for changes in structure in dynamic (i.e., temporal) networks over time. Such an approach may also be applied to the the visual analysis of misrepresentations in structure, such as misrepresentative accounts and their associated activities.

\subsection{XAI to understand the AI behind misrepresentation}

Using XAI (eXplainable AI) in misrepresentation detection is important to address the need for transparency in automated systems that determine the veracity of news content. The critical advantage of XAI in this context is its ability to uncover biases and provide insights into the decision-making process of detection models. The XAI features for low quality news detection include user behavior, content authenticity, and stylistic elements of the written content, which are processed by machine learning models with explainability layers or post-hoc interpretation methods to make decisions transparent \cite{yang2019xfake, shu2019defend, lu-li-2020-gcan, kurasinski2020towards, silva2021propagation2vec, xiangyu2023relevance}.

The authors of \cite{lu-li-2020-gcan} proposed a low quality news detection model called Graph-aware Co-Attention Networks (GCAN). This model can determine whether a short-text tweet is fake, given the sequence of its retweeters, and generate explanations by emphasizing evidence from suspicious retweeters. The work by \cite{szczepanski2021new} developed two XAI techniques, Local Interpretable Model-Agnostic Explanations (LIME) and Anchors, to evaluate low quality news data. The authors of \cite{xiangyu2023relevance} proposed a layer-wise relevance propagation (LRP) based method as a post-hoc explanation to observe how the neighboring nodes affect the low quality news prediction results for a node to explain. The work by \cite{yong2023explainability} used Naive Bayes, Support Vector Machines (SVMs), Logistic Regression, Decision Tree, and Random Forest to explore data, and explain the decision-making process of their AI systems for detecting COVID-19 Twitter low quality news.

Explainability in misrepresentation detection in image analysis is also a growing area of interest, as the lack of transparency in deep learning models limits their trustworthiness and adoption. Current methods often rely on ``black-box'' classifiers, which provide little insight into how decisions are made by default, so researchers have begun to explore XAI techniques for fake image detection. For instance, gradient-based methods, such as Grad-CAM \cite{selvaraju2017grad}, have been used to visualize the regions of an image that contribute most to the detection decision. Similarly, attention mechanisms \cite{vaswani2017attention, chefer2021transformer} in transformer models can highlight the most relevant features for distinguishing real from fake content. The work of Yang et al. \cite{yang2024snippet} assessing explanations from a subjective aspect has shown that these methods can improve interpretability, but their effectiveness depends on the quality and diversity of the training data.

\subsubsection{Machine learning explainability visualization}
\label{sec:explainabilityVis}

Data visualization is also often used as a tool to support XAI of many of the ML models used by many of the techniques described in Section~\ref{sec:contentDimension} and in Section~\ref{sec:actorDimension}. Visual analytics have been recognized as a means by which  machine learning can be explained and trusted.\cite{Chatzimparmpas2025}. ML models can be supported by data visualizations to enable users to comprehend the reasoning behind decisions (explain-ability) and the inner workings of how a specific decision is made (interpretability). Explainability is not simply a visualization of the ML feature space but maps from the feature space to entities and terms that a non-expert user can recognize (translate from feature-related terms to human-understandable terms, i.e., feature-space to human-space). For example, Yang et al. \cite{yang2019xfake} presented the XFake system that assists end-users in identifying news credibility by using different components to jointly consider both attributes and statements of news, and includes visualization as part of the system, to better understand the outputs of the underlying ML frameworks.

Histograms are used to visualize numeric attributes such as attribute significance and prediction score. They also use heat maps to highlight the importance  of words and phrases in the output, and employ ensemble trees to show the overall structure of the decision process, as well as the activated paths for a specific decision. Ensemble visualizations have indeed been used for ML interpretability for some time \cite{EnsembleMatrix2009},  and are just one of many approaches available to help make ML approaches more trustworthy. Other visualizations for explainability adopt standard scatter plot approaches, as described by \cite{Chatzimparmpas2020}. However, in recent years there has been significant and rapid progress and research into the use of visualization for model interpretability, known as VIS4ML, is a fast-growing field \cite{Wang2024}.

\section{Practicalities of platform regulation, data access, and interventions}
\label{sec:discussion}

In this final section, we discuss topics that affect the ability to research, identify, and address strategic misrepresentations on OSNs, as well as some of the consequences. As our approach is grounded in a pragmatic, data-driven four-dimensional model, attention to issues of data access, regulatory frameworks, and established interventions within the online social system constitute a relevant endeavor. The pragmatism underpinning the proposed model entails that, while regulations and interventions inevitably carry a normative dimension, our primary emphasis lies on their practical implications in terms of a bottom-up, contextually-sensitive identification of strategic misrepresentation. On this basis, we discuss the potential effects of recent regulatory measures and commonly deployed interventions within complex social systems characterized by substantial levels of indeterminacy. It should be noted, however, that both our model and the approaches outlined above require access to multiple types of data that are often very sensitive, including data related to malicious actors and high-risk topics, as well as data from ordinary users, in order to establish counterfactual benchmarks and estimate what we have termed baseline representations.

\subsection{Regulation of online platforms}

Social media companies, also known as Very Large Online Platforms (VLOPs), monetize users' data and attention while exerting significant social influence. They collect personal and inferred data to build user profiles for targeted advertising \cite{RN45}, and use algorithms to optimize engagement, shaping users’ information exposure and the content most likely to go viral \cite{RN235}.

VLOPs like Google, X, Meta, and TikTok exert an outsized influence on society, culture and politics by creating  multi-functional digital ecosystems in which users can connect with their network, keep up with news, and even buy and sell products. Such companies exert their ``digital dominance'' in the market through practices like excessive data collection, tying or bundling services, acquisition of challengers and competitors, and restricting rivals' platform access. This market dominance of VLOPs creates challenges for users in terms of choice, trust, and access to services. The business model of social media functions through maximizing user engagement, therefore design decisions made by their developers align with this goal. Indeed, bias in platform design is evidenced by Facebook's implementation of the Meaningful Social Interactions (MSI) algorithm in 2018, which was intended to stem the spread of disinformation but instead increased it \cite{RN235}.

Data is also exploited externally to the VLOPs themselves. In 2018 the Cambridge Analytica affair revealed that users' personal data exposed on Facebook was sold to the consulting firm Cambridge Analytica and subsequently used by political actors to influence election campaigns in the US and UK  \cite{theguardianMadeSteve}. While the resulting controversy led to some reforms \cite{RN300}, it had little impact on the development of data brokering as a pillar of the social media business model \cite{RN275}. As the Cambridge Analytica affair highlighted, this data is immensely valuable not only to social media companies and advertisers but also to those who wish to exert political influence. Recently, there have been efforts to regulate VLOPs through legislation, the most prominent example of which is the Digital Services Act of the European Commission. 

\subsubsection{European Union's approach}

The Digital Services Act (DSA) became EU law in November 2022, applying in full in all EU states from February 2024. It aims to create a safer digital space by setting clear responsibilities for online platforms to tackle illegal content, protect users’ rights, and ensure transparency. The DSA mandates that VLOPs implement robust content moderation systems, conduct risk assessments, and provide transparency reports. This legislation aims at reducing the spread of misinformation by holding platforms accountable and ensuring they take proactive measures to mitigate risks. The DSA also emphasizes the importance of user empowerment and protection. By requiring platforms to provide users with tools to report harmful content and appeal moderation decisions, the DSA aims to foster a more transparent and accountable online environment~\cite{europaImpactDigital}.

Self-governance by VLOPs has proven to be inadequate in addressing online misbehavior and strategic misrepresentation. Platforms have been criticized for inconsistent enforcement of their policies, lack of transparency, and prioritizing profit over public safety. Moreover, self-governance often lacks the necessary oversight and accountability. Without external regulation, platforms may not have sufficient incentives to invest in effective content moderation and risk mitigation strategies. Within the DSA framework, the 2022 strengthened code of practice on disinformation (COPD) slightly minimizes an over-reliance on self-regulation of previous EU legislation; however, important issues remain. Essentially, self-governance is still the main approach to content moderation, algorithmic accountability, advertisement, and recommender systems \cite{nannini_beyond_2025}, all aspects crucial to monitor how strategic misrepresentations can be spread. Furthermore, the COPD is not a binding law. Indeed, X (formerly Twitter) is not a signatory of COPD 2022 despite being identified as the primary platform for the dissemination of misrepresentations \cite{RN299}. The European Commission has already opened proceedings against several VLOPs for potential breaches of the DSA. In the proceedings against X, preliminary findings have been released, identifying evidence of dark patterns, lack of advertising transparency, and lack of data access for researchers~\cite{europaPressCorner}.

\subsection{Researcher access to data}

Access to social media data for researchers is essential to investigate platforms practices and user behavior. Different platforms have their own procedures. The Bluesky platform provides access through an open API,  while others such as TikTok provide researcher tools or APIs that researchers need to apply for access to. Free research access to X is restricted to cases specified under the Digital Services act, data must be paid for otherwise.
 
In article 40 of the DSA, it states that vetted researchers will be able to ``request data from very large online platforms (VLOPs) and search engines (VLOSEs) to conduct research on systemic risks in the EU'' \cite{eudigitalservicesactArticleDigital}. As of September 2025, the DSA platform is live and available for vetted researchers to make requests for data sets \cite{DSA2025}. Ensuring the access of independent research teams to data is particularly important given that previous research has warned that the level of independence of auditors when analyzing compliance with DSA of VLOPs remains unclear \cite{nannini_beyond_2025}. Feasible data access thus has the potential to contribute to big platforms' accountability and to uncover breaches of the DSA. Additionally, data access for research purposes can enable researchers to analyze, identify, and document examples of strategic misrepresentation that may not be considered disinformation within the DSA framework, thereby providing on-the-ground examples \cite{goanta_great_2025} that can ultimately help improve current legal frameworks aimed at fostering safe online environments.

However, researchers' access to data, and their capacity to effectively use it, cannot be understood only through the DSA framework. Another important layer to consider is GDPR (General Data Protection Regulation). GDPR is an EU law that took effect in 2018, with the aim of regulating how public and private organizations collect, process, store, share, and communicate European citizens' personal data. Its objective is to protect individuals' right to privacy, so that they have the agency to give, deny, or withdraw their consent to organizations collecting and processing their data.

With regard to the multidimensional model that we propose to detect multilayered evidence as indicators signaling the occurrence of strategic misrepresentation, while the DSA makes it easier for researchers to access data, GDPR impacts and potentially limits or slows down different tasks that compose the research process. Without aiming to be exhaustive, previous research \cite{greene_adjusting_2019} has shown that GDPR compliance, among other things, makes it difficult to generate  longitudinal datasets and imposes strict conditions on processing sensitive data (including data on socio-demographics, political beliefs, and opinions). Researchers, therefore, need to spend time justifying why data processing is carried out in the public interest.

Additionally, and this is especially important for the pragmatic, bottom-up approach to data that our model proposes, the GDPR’s principle of data minimization implies that researchers need to collect ``only what is strictly necessary'' for their objectives. This can be difficult to assess for models, such as ours, which are meant to flexibly capture trends and dynamics emerging from data related to content, actors (i.e., accounts), and processes (i.e., between-account interactions). Calculating the `right and minimal' amount of data in data-driven research is oftentimes very challenging.

Overall, tensions exist between the DSA and the GDPR frameworks, creating a paradox: while the DSA allows for easier data acquisition, GDPR poses different challenges for effective and timely data processing for research purposes in matters that may certainly be of public interest.

\subsection{Notice-and-takedown procedures}

Through notice-and-takedown procedures, the DSA focuses on the usual approach towards the presence of misrepresentation online: post-hoc content corrections and debunking \cite{nannini_beyond_2025}. However, as we highlighted throughout the text, misrepresentation constitutes an interactional process within OSNs that transcends content and epistemics as its only dimension  \cite{hayward_problem_2025}. Analyses of strategic misrepresentation should attend to epistemic issues, but also consider the properties of accounts, the covertness of online activities, and the various processes underpinning multi-account coordination. As we underlined in the introduction, reality and ``truth'' is also an outcome of human interaction. This implies that specific information, beliefs, and values take part in both group formation and group identity-construction processes \cite{bliuc_opinion-based_2007}.

On the technical side, this implies the need for multidisciplinary approaches that analyze the content pushed by different forms of online misbehavior, as well as the structural relationship between the accounts involved and their characteristics, while also paying attention to the particular contexts in which misrepresentations are being spread \cite{goanta_great_2025}. Along these lines, interventions should move beyond post-hoc fact-checks. We suggest that showing citizens their position within the broader online information space can raise awareness of how platform affordances, personal location in that space, and coordinated action (automatized, organic, or mixed) interact to produce misrepresentation and shape its effectiveness.         

\subsection{Outcomes of interventions and deplatforming to address strategic misrepresentation}

The shattering of the ideal of social media as platforms where a more bottom-up, inclusive, and transparent public sphere stemmed from the networks of outrage and hope created by social movements that wanted to expand democratic rights \cite{castells_networks_2012}, has given way to debates about social media content regulation. Nowadays, the growing presence of different forms of online misbehavior, and their potential psychological, social, and political consequences also underline the need for platforms’ accountability and regulation.  The majority of users are in favor of some sort of content regulation, particularly threatening content and content inciting violence. However, who should regulate content (the platforms, the government or users themselves), what are the limits of this regulation, and what is the abstract value of ``freedom of speech'', are contested issues. This is the conclusion of a report from the Technical University of Munich and the University of Oxford that surveyed people’s opinions about these topics in 10 countries \cite{theocharis_public_2025}.

Different types of regulations from governments, supranational institutions (e.g., the EU), and platforms were and are still being discussed. Here, we develop some general principles that were discussed throughout the paper, and apply them to the context of potential interventions in the face of strategic misrepresentation. The latter provides tools for users and institutions to foster our societies' resistance against coordinated and inauthentic attempts to skew public opinion.  

Interventions should consider the online social system as a whole, as a dense network of interconnected accounts that sometimes exploit both the algorithmic preferences and the platform's affordances, as well as the audiences' leanings and group biases, to make a particular message or narrative viral. One aspect that makes online misbehavior consequential is coordination between accounts and the strategic use of information. Bearing these aspects in mind implies a shift from unidimensional approaches (e.g., account-centered approaches) to identify and potentially combat strategic misrepresentation, to multidimensional approaches focused on content, actors, and processes. Strategic misrepresentation is often enabled through covert tactics involving interactions between accounts, which are likely to produce misrepresented versions of the online social system and/or the different social groups that take part in the latter. The following insights are useful to generate general principles to promote the most effective interventions possible in online environments:

\begin{enumerate}
    
    \item An intervention (e.g., policies, regulations) needs to consider not only what the effects are going to be in a particular part of the online social system, but also on the whole. The online social system is a complex system in which many constitutive parts are constantly interacting among themselves and with the ``outside world''. Thus, a small regulation intending to impact any of these constitutive parts might also exert an impact on the whole. 
    
    \item Deplatforming specific influential figures that spread misrepresentations is an intervention with potential limitations. While it can decrease their overall audience (at least temporarily), there is a risk that these banned users will migrate to other platforms where they may become more active and spread even more toxic content. \cite{ali_understanding_2021}. In addition, there is ample evidence that some sociopolitical groups succeed in gaining social support by mobilizing dominant-group victimhood \cite{reicher_resentment_2020}. Some interventions, including high-profile deplatforming cases, can thus be framed as censorship of group-relevant beliefs and ideas, leading to identity construction processes in which certain social sectors see themselves as victims of unfair and politically biased actions. Furthermore, given that virality is a collaborative accomplishment among various social actors, the prevalence and social-psychological impact of misrepresentative narratives that have become widely popular and group-relevant is unlikely to diminish simply by deplatforming one or a few of their promoters. Consequently, decisions around regulation, censorship, and deplatforming must be based on context-sensitive assessments of their potential effects within the complex, networked character of online social systems.
    
    \item Interventions in the online social information systems can have unintended consequences: This is because we are dealing with complex, adaptive systems characterized by high levels of interdependence, non-linearity, and indeterminacy. Actions aimed at altering the behavior of specific agents or accounts (e.g., banning users or removing content) can trigger emergent dynamics that are difficult to predict. We are dealing with dynamic multi-causal processes with different plausible trajectories \cite{durrheim_human_2025}, so small interventions can lead to cascading effects across the network, with the capacity even to produce a big impact on the system as a whole \cite{quayle_social_2025}. This calls for context-sensitive, system-aware approaches to platform governance that account for the dynamic, multifaceted, and multi-agent nature of influence online.
    
\end{enumerate}

\section{Conclusion}
\label{sec:conclusion}

In this article, we have argued that debates on online disinformation must move beyond a narrow focus on false content alone. Although disinformation is typically distinguished from misinformation by its intentional dissemination of veridically false material, information campaigning in online social networks (OSNs) operates as much through the manipulation of social structures as through the fabrication of content. Coordinated authentic users, incentivized human agents and automated accounts can collectively distort visibility, popularity and perceived consensus, thereby shaping meaning at scale. We have therefore reframed the problem as one of strategic misrepresentation -- that is, any \textit{additive distortion to the baseline representation} arising from the intentional manipulation of content, actors or processes within and across OSNs. Addressing this phenomenon demands a multidisciplinary, pragmatist approach capable of capturing how such misrepresentation is collectively accomplished in practice and of informing more precise analytical and policy responses.

We contend that such pragmatist reframing is necessary to move beyond the assumption that reality is self-evident, and that social information can be simply classified as true or false by how well it mirrors the independent properties of the world. Our reframing accounts for the fact that humans are epistemically interdependent, and that what is regarded as ``truth'' is often the evolving outcome of collective meaning-making processes. Our reframing does not just focus on the content and whether what is being said is ``true'' but rather whether who is saying it is genuine, and whether there is a distortion to the process by which a signal disseminates either within or across OSNs. This often includes manipulating account entities within the network and leveraging the network itself through a degree of non-trivial coordination. Strategic misrepresentation cannot disseminate through the actions of individuals in isolation, and addressing it therefore requires analysis of the interactions among multiple accounts as well as an understanding of the emergent properties of OSNs. 

Our framework defines and classifies strategic misrepresentation via four key dimensions: content distortion, actor distortion, process distortion and covertness, which we respectively operationalize through the concepts of epistemic diligence, behavioral authenticity, non-trivial coordination and transparent disclosure. However, applying the framework is more than a simple ticking off of dimensions. The task of a researcher, or anyone applying our framework, is to triangulate evidence from various sources to take a holistic view that encompasses the three first dimensions, each constituting a particular deviation from the baseline representation. Once a deviation is found, the fourth dimension, covertness, then calls for a more qualitative evaluation to rule out any legitimate reason underlying misrepresentation in at least one of the other dimensions, be it native political advertising, a promotional campaign, or an underground democratic movement operating under a repressive regime. For example, in the absence of actor and process distortion, a campaign focusing exclusively on a politician's achievements probably lacks epistemic diligence, but is also immediately recognizable as a legitimate political campaign (i.e., subject to local context) and therefore fulfills the criteria on transparent disclosure.

Another edge case is when signal amplification is achieved not by distorting actors or processes but by selecting messages that are gaining most traction in the baseline. Consider the example of an interest group trying to publicize an issue. The organization might frame the issue in several ways and coordinate the dissemination of the resulting messages through different people before selecting the most promising framing in terms of virality metrics, to publicize it further \cite{zeitzoff_how_2017}. Such an activity is not necessarily overt and requires a degree of coordination, but as long as the content is epistemically diligent and does not involve actor or process distortion, it is unlikely to qualify as strategic misrepresentation. However, if there is a spectator standing on the opposite side of the issue, such message tailoring is likely to resemble strategic misrepresentation in spirit. Caveats aside, we emphasize that the methodology underpinning our framework and its operationalization of strategic misrepresentation is explicitly pragmatist in orientation. Accordingly, it cares mostly about how phenomena manifest in practice, rather than about advancing a strong normative position on how they ought to be.

To support the adoption of the strategic misrepresentation framework as a lens on the contemporary hybrid media system, we have surveyed state-of-the-art techniques encompassing machine learning and network science to detect content, actor and process misrepresentation. We have proposed a unified approach to detection that rests on the notion of multilayer networks, with data across the three dimensions populating the vertices, edges and layers of a unified graph structure. As a concrete tool for practitioners, we have also described key visual analytics approaches to grasp strategic misrepresentation and understand the functioning of the various detection techniques across layers. We have also noted how addressing strategic misrepresentation is not solely a technical problem, but involves a number of practical constraints. We have discussed for example the regulation of OSNs and the challenge of accessing their data in a meaningful way. We envision our work will provide researchers, practitioners and decision-makers with a novel perspective on identifying strategic misrepresentation in OSNs, all while equipping researchers with the most relevant and applicable tools to address the normatively most concerning aspects of strategic misrepresentation.

Looking ahead, several research directions open up from the unified conceptualization and detection of strategic misrepresentation in OSNs. First, do isolated forms of misrepresentation interact in ways that produce amplification effects or other emergent phenomena which would be more, and different, than the sum of its parts? Such systemic distortions which cannot be reduced to their individual components would have implications to how robust or vulnerable online platforms are against strategic misrepresentation, and how these challenges should be addressed. Second, does combining different forms of misrepresentation alleviate the problem of increasing difficulty in detecting misrepresentation as compared to methods that do not consider the context of isolated misrepresentation instances? There are also questions about whether misrepresentation manifests in different ways on different platforms, tailored on the technical features and interaction paradigm of the platform \cite{cinelli_echo_2021}. Given the current trend in deplatforming, this can lead to settings where similar phenomena are replicated over multiple platforms, and different algorithmic solutions and social ecosystems are tested against each other revealing which ones are more robust against various types of misrepresentation. Addressing these questions will require methodological innovation and interdisciplinary collaboration.

\section{Acknowledgments}
\label{sec:acknowledgements}

The authors would like to acknowledge the support of CHIST-ERA for the CON-NET project and their local funding agencies for this grant. The agencies are the Fonds National de la Recherche Luxembourg (grant: INTER/CHIST22/17002518/CON-NET), the Irish Research Council (grant: Chist ERA-CW68565), the Scientific and Technological Research Council of Türkiye (grant: 222N311), the Research Council of Finland (grant: 357743) and the Onderzoeksprogramma Artificiele Intelligentie (AI) Vlaanderen programme. Also funded in part by the European Research Council (ID-COMPRESSION, grant: 101124175).

Funded in part by the European Union. Views and opinions expressed are however those of the authors only and do not necessarily reflect those of the European Union or the European Research Council Executive Agency. Neither the European Union nor the granting authority can be held responsible for them.

\bibliographystyle{acmreferenceformat}
\bibliography{con_net_survey}

\end{document}